\documentclass[twocolumn]{aastex63}
\usepackage[utf8]{inputenc}
\usepackage{graphicx}
\DeclareGraphicsExtensions{.ps,.eps,.pdf,.jpg,.png}

\newcommand{\kms}{\ifmmode{\rm km\thinspace s^{-1}}\else km\thinspace s$^{-1}$\fi}
\newcommand{\hd}{HD~149834}

\begin{document}


\title{Discovery and Characterization of a Rare Magnetic Hybrid $\beta$~Cephei Slowly Pulsating B-type Star in an Eclipsing Binary in the Young Open Cluster NGC~6193}
\author[0000-0002-3481-9052]{Keivan G.\ Stassun}
\affiliation{Department of Physics and Astronomy, Vanderbilt University, Nashville, TN 37235, USA}
\author[0000-0002-5286-0251]{Guillermo Torres}
\affiliation{Center for Astrophysics $\vert$ Harvard \& Smithsonian, 60 Garden Street, Cambridge, MA 02148, USA}
\author[0000-0002-3054-4135]{Cole Johnston}\affiliation{Instituut voor Sterrenkunde, KU Leuven, Celestijnenlaan 200D, 3001 Leuven, Belgium}
\affiliation{Department of Astrophysics, IMAPP, Radboud University Nijmegen, P. O. Box 9010, 6500 GL Nijmegen, the Netherlands}
\author[0000-0002-5951-8328]{Daniel J. Stevens}
\affiliation{Department of Astronomy \& Astrophysics, The Pennsylvania State University, 525 Davey Lab, University Park, PA 16802, USA}\affiliation{Center for Exoplanets and Habitable Worlds, The Pennsylvania State University, 525 Davey Lab, University Park, PA 16802, USA}
\author[0000-0002-2457-7889]{Dax L.\ Feliz}
\affiliation{Department of Physics and Astronomy, Vanderbilt University, Nashville, TN 37235, USA}
\author[0000-0002-5365-1267]{Marina Kounkel}
\affiliation{Department of Physics and Astronomy, Vanderbilt University, Nashville, TN 37235, USA}
\author[0000-0002-0514-5538]{Luke G.\ Bouma}
\affiliation{Department of Astrophysical Sciences, Princeton University, 4 Ivy Lane, Princeton, NJ 08540}

\begin{abstract}
As many as 10\% of OB-type stars have global magnetic fields, which is surprising given their internal structure is radiative near the surface. A direct probe of internal structure is pulsations, and some OB-type stars exhibit pressure modes ($\beta$~Cep pulsators) or gravity modes (slowly pulsating B-type stars; SPBs); a few rare cases of hybrid $\beta$~Cep/SPBs occupy a narrow instability strip in the H-R diagram. The most precise fundamental properties of stars are obtained from eclipsing binaries (EBs), and those in clusters with known ages and metallicities provide the most stringent constraints on theory. Here we report the discovery that HD~149834 in the $\sim$5~Myr cluster NGC~6193 is an EB comprising a hybrid $\beta$~Cep/SPB pulsator and a highly irradiated low-mass companion. We determine the masses, radii, and temperatures of both stars; the {$\sim$9.7}~M$_\odot$ primary resides in the instability strip where hybrid pulsations are theoretically predicted. The presence of both SPB and $\beta$~Cep pulsations indicates that the system has a near-solar metallicity, and is in the second half of the main-sequence lifetime. The radius of the {$\sim$1.2}~M$_\odot$ companion is consistent with theoretical pre-main-sequence isochrones at 5~Myr, but its temperature is much higher than expected, perhaps due to irradiation by the primary. The radius of the primary is larger than expected, unless its metallicity is super-solar. Finally, the light curve shows residual modulation consistent with the rotation of the primary, and {\it Chandra} observations reveal a flare, both of which suggest the presence of starspots and thus magnetism on the primary. 
\end{abstract}

\section{Introduction}
Stellar clusters and associations offer the opportunity to test and refine theories 
of stellar structure and evolution. Given their common distance, chemical 
composition, and assumed common age, clusters act as laboratories to calibrate 
models and physical mechanisms that must be able to simultaneously explain
the observed characteristics of all cluster members. Individual stars which 
exhibit some form of variability, such as rotational modulation, wind variability, 
stellar pulsations, and/or eclipses offer an additional opportunity to derive 
more precise stellar information once modelled. Thus, the identification of
stars which exhibit one or more forms of variability is important for the 
advancement of the theory of stellar structure and evolution. 

OB type stars are observed to have high projected rotational velocities and are
expected to rotate quickly at birth \citep{Huang:2006a,Daflon:2007,Garmany:2015}. Some 20\%
of B stars are observed to exhibit emission in their line profiles and are classed 
as Be stars \citep{Rivinius:2013}, a phenomenon which is associated with near-critical 
rotation \citep{Porter:2003}, non-radial pulsations \citep{Semaan:2018}, and/or
binarity \citep{Bodensteiner:2020}. In the absence of the transient Be phenomenon, 
B stars are expected to only show rotational modulation in the event that
surface features are introduced by magnetic fields or chemical inhomogeneities. 
However, only some 10\% of OB stars are observed to have global magnetic fields, 
making this phenomenon rare and difficult to explain theoretically \citep{Wade:2019}. 
The incidence of stars with a detected global surface magnetic field in close 
binaries ($P_{orb}<10$~d) is even smaller \citep[$<2$\%;][]{Alecian:2015, 
Neiner:2015}, although chemically peculiar stars such as HgMn stars are observed 
to have a very high binary fraction, despite null magnetic detections 
\citep{Takeda:2019}. Alternatively, gyrosynchrotron emission has been detected 
in several OB binaries, implying a magnetic field \citep{Gagne:2011}.

There is a wide variety of observed pulsational variability on the upper main-sequence. 
OB stars are observed to exhibit either pressure (p) modes as $\beta$~Cep pulsators, or gravity 
(g) mode pulsations as slowly pulsating B-type stars (SPBs) driven by the $\kappa$-mechanism 
associated with the Fe-group opacity bump \citep{Moskalik:1992,Dziembowski:1993,Pamyatnykh:1999}. 
Additionally, late O to early B type stars are observed to exhibit both p and g mode pulsations 
simultaneously as hybrid pulsators, such as $\gamma$ Peg \citep{Handler:2009}. Theoretical 
instability strips for OB stars are highly sensitive to rotation, metallicity, and opacity tables 
\citep{Salmon:2012,Moravveji:2016,Szewczuk:2017}, and often have trouble reproducing the 
total range over which pulsations are observed. With the advent of space missions, OB 
pulsators are being observed with an unprecedented duty-cycle, revealing an even larger 
diversity of pulsators that theoretical instability strips struggle to explain
\citep{Pedersen:2019,Balona:2020,Burssens:2020,Labadie-Bartz:2020}. 
Even in the absence of detailed modelling, the presence of pulsations in OB stars reveals 
a nearly Solar metallicity in order to excite modes via the $\kappa$-mechanism. 

{Adding to the diversity of variability, massive OB stars are also host to 
internal gravity waves (IGWs), which are generated by the turbulent motions at 
the boundary of the convective core and the radiative envelope. Such waves are 
theoretically predicted by simulations to produce observable photometric variability 
\citep{Rogers:2013,Edelmann:2019} at low frequencies. The observed low-frequency 
excess in a large sample of CoRoT, Kepler, and TESS targets has been interpreted
as being caused by IGWs \citep{Blomme:2011,Bowman:2019b,Bowman:2020a}. As the 
presence of IGWs is not predicted or yet observed to be dependent on metallicity, 
it offers an explanation for the stochastic variability often observed in massive
stars that cannot be explained by sub-surface convection \citep{Cantiello:2009,
Bowman:2019b}.}

The most precise estimates of stellar parameters are obtained via eclipsing binary star systems, which can produce estimates on stellar
masses and radii to the percent level \citep{Torres:2010}. When compared against
grids of stellar models or isochrones, the age, core-mass, and other internal 
properties can be inferred to high precision, providing excellent evolutionary 
diagnostics. 

Similarly, clusters can be fit with such isochrones in order to determine
the age or turn-off mass of the cluster. However, clusters have been observed
to exhibit features such as split main-sequences, extended main-sequence turn-offs (eMSTO), 
and complex chemical profiles which challenge the assumption that clusters are
constituted of a single population of stars. Studies have demonstrated, however, 
that blue stragglers, and other split main-sequences can in part be explained by 
a single population, some percentage of which has undergone binary interactions, 
diverting stars from an otherwise apparently single star evolutionary trajectory \citep{Beasor:2019}.
Additionally, rapid rotation and its effects have been invoked to explain the extended
main-sequence turn-off phenomenon through either photometric color changes caused by 
the von Zeipel effect and/or enhanced chemical mixing induced by rotational shears 
\citep{Bastian:2016,Dupree:2017,Kamann:2018,Marino:2018}. Alternatively, it has been 
shown the eMSTO phenomenon can be explained at least in part by a single population 
of stars with varying overshooting and envelope mixing efficiencies \citep{Yang:2017,
Johnston:2019}. 

However, there is not a single mechanism which can explain the 
diversity of observations of young massive clusters \citep{Bastian:2017}. Instead,
many questions still remain about whether the process of star formation in clusters
is a rapid process that occurs via fragmentation within two times the free fall timescale
of the cluster, or if it is instead a longer process that potentially spans millions 
of years \citep{Getman:2014}. Recently, observations have revealed that stars in 
the cores of young clusters are younger than stars in the halo, suggesting an outside-in
formation scenario \citep{Getman:2014}. Further corroboration of this scenario can 
be sought through the modelling of clusters which have individual stars that exhibit 
multiple forms of variability. To this end, clusters with several young, massive, pulsating 
or rotationally variable stars in eclipsing binaries offer the best means of scrutinising 
formation scenarios with multiple independent constraints.

In this paper, we characterise the B2V + K/MV binary HD~149834 containing an SPB / $\beta$ Cep 
pulsating primary. Using archival radial velocities and photometry combined with new space-based 
photometry from the Transiting Exoplanet Survey Satellite \citep[TESS,][]{Ricker:2015} we perform 
eclipse and spectral energy distribution modelling of the system and investigate the residuals for 
asteroseismic and rotational signatures. Furthermore, we use the parallax and space motions from Gaia 
DR2 \citep{Lindegren:2018,Luri:2018} to investigate the membership probability of this system. Finally, 
we discuss this system in an evolutionary context and conclude in Sections \ref{sec:discussion} and 
\ref{sec:summary}.

\section{The HD~149834 Binary System: Age, Cluster Membership, and Distance Considerations}\label{sec:hd149834} 

\begin{figure}[!ht]
    \centering
    \includegraphics[width=\linewidth,trim=40 180 50 140]{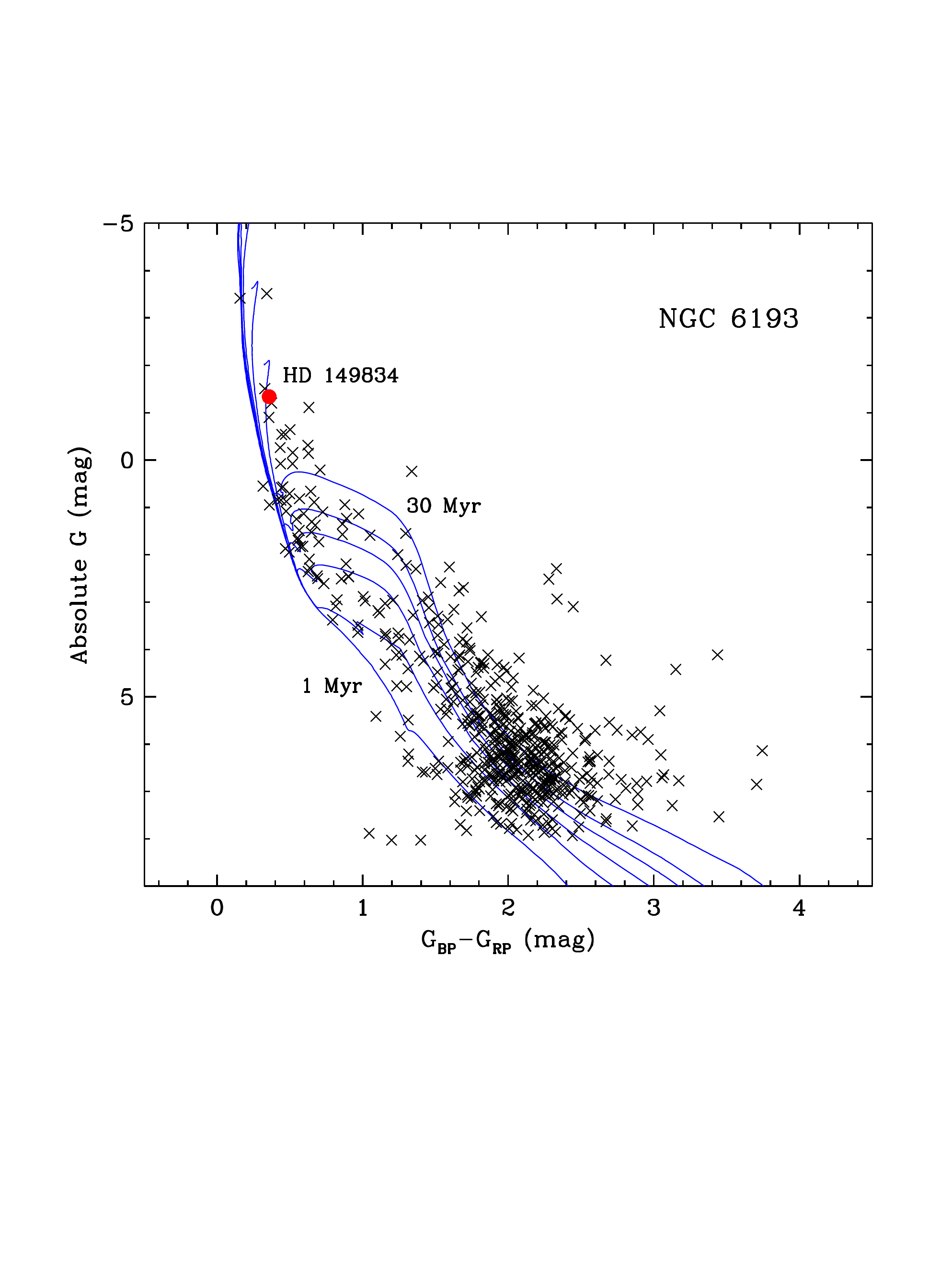}
    \caption{CMD of the parent cluster. The oldest of the MIST isochrones shown (1, 2, 3, 5, 10, 30 Myr) is the one that comes closest to the location of \hd.}
    \label{fig:cmd}
\end{figure}

HD~149834 was identified as a single-lined spectroscopic binary (SB)
by \citet{Arnal:1988} in their study of likely members of the young cluster NGC~6193. 
%
More recently, HD~149834 is listed among the $\sim$550 stars considered by \citet{Cantat-Gaudin:2018} as probable members of NGC~6193 (defined as having a probability of membership of 0.5 or above), based on {\it Gaia\/} DR2 parallaxes and proper motions. However, HD~149834 itself just barely satisfies their membership criteria, having a probability of 0.5. The {\it Gaia\/} color-magnitude diagram (CMD) of these stars (Figure~\ref{fig:cmd}) suggests an age of a few Myr, consistent with previous estimates in the literature, whereas the position of HD~149834 in the CMD is more consistent with an older age of $\sim$30~Myr. 

As we will show in Section~\ref{subsec:lcfit}, the radial velocities reported by \citet{Arnal:1988}, refit by us along with the TESS photometry, yield a systemic velocity of $-{34.7 \pm 1.8~\kms}$, which agrees well with the average for NGC~6193 of $-34.4 \pm 2.0~\kms$ from all 18 stars observed by those authors, or $-33.2 \pm 1.1~\kms$ from just the six non-variable stars in \cite{Arnal:1988}. This lends additional confidence that it is a true member. 

\begin{figure*}[!ht]
    \centering
    \includegraphics[width=0.49\linewidth]{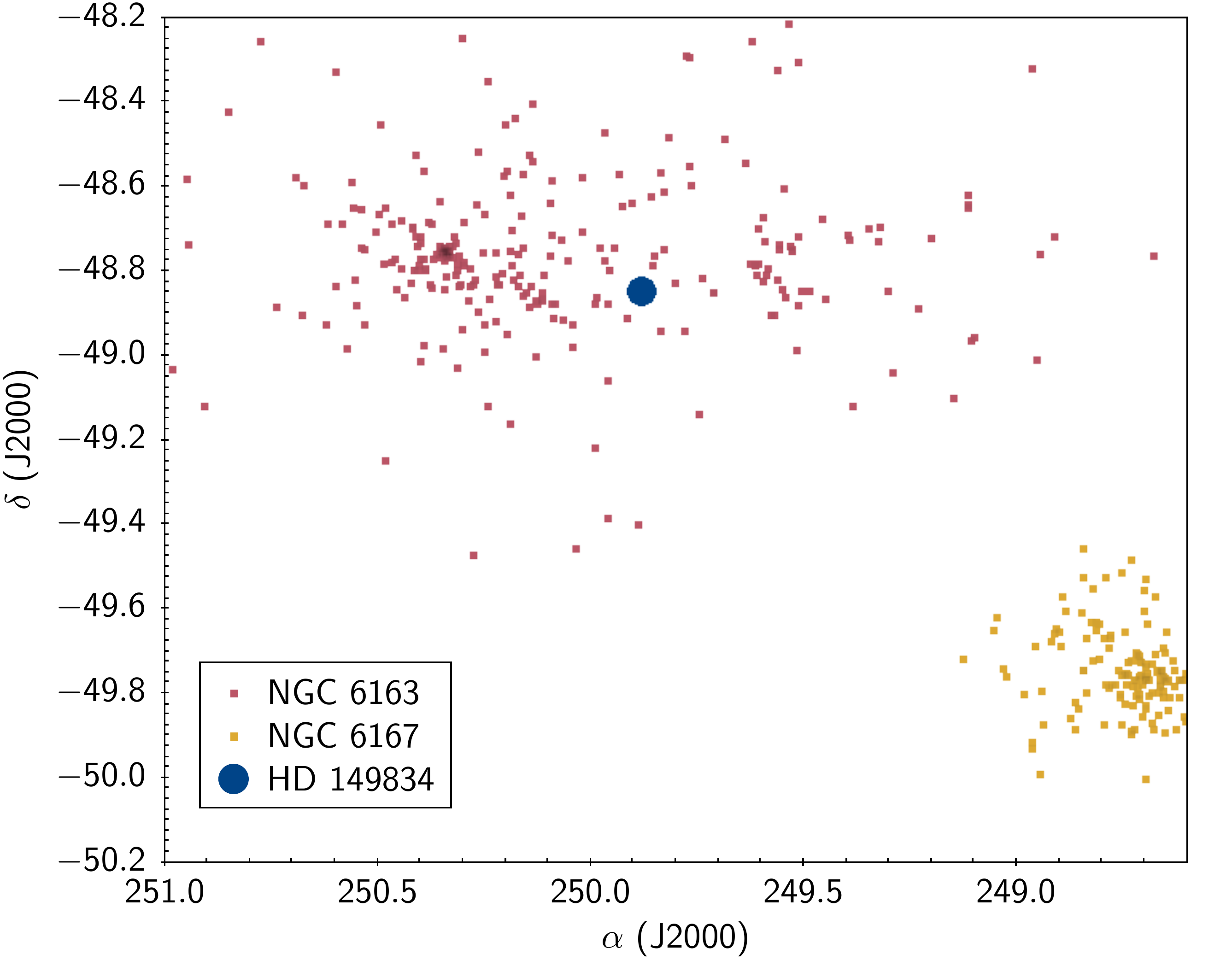}
    \includegraphics[width=0.49\linewidth]{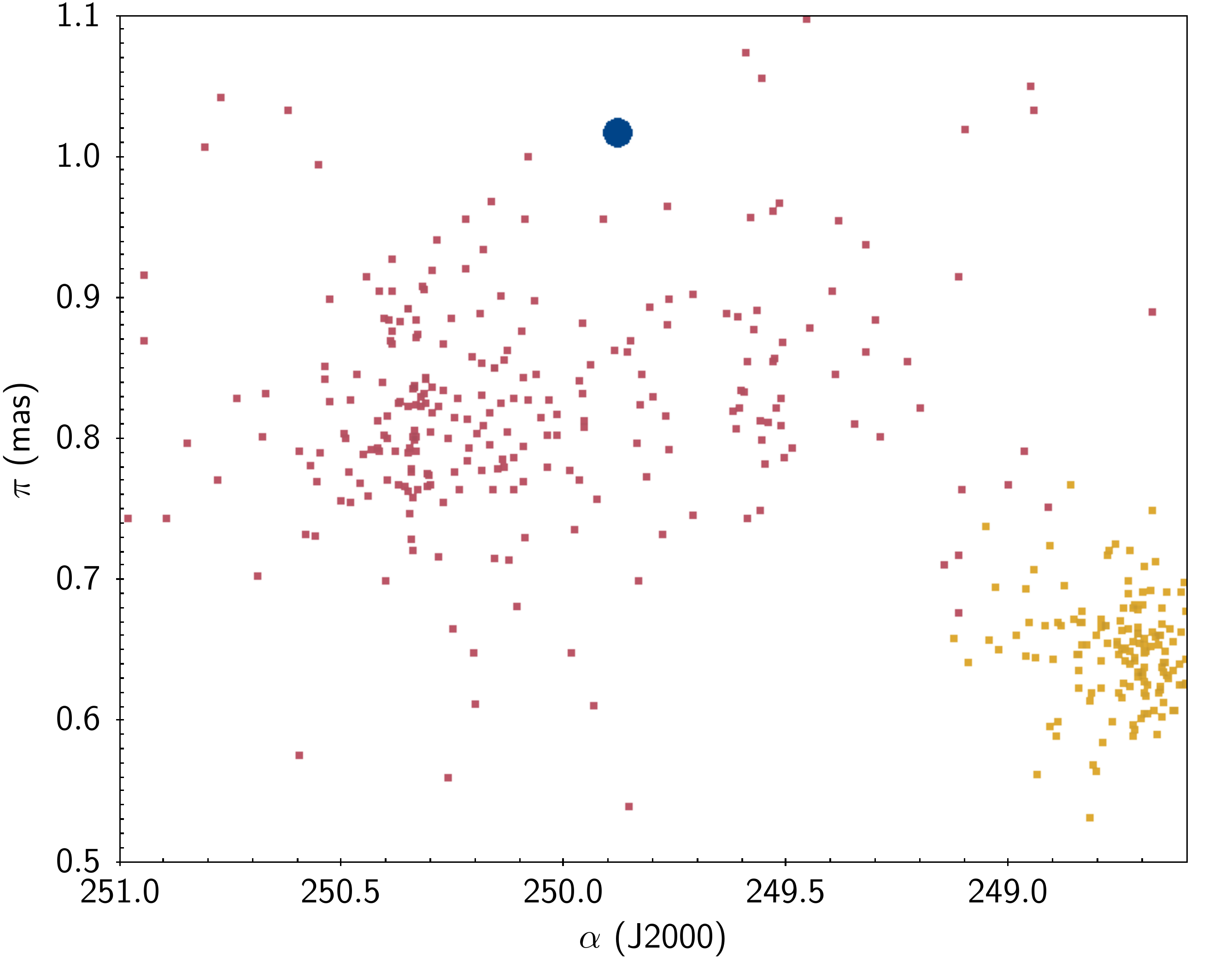}
    \includegraphics[width=0.49\linewidth]{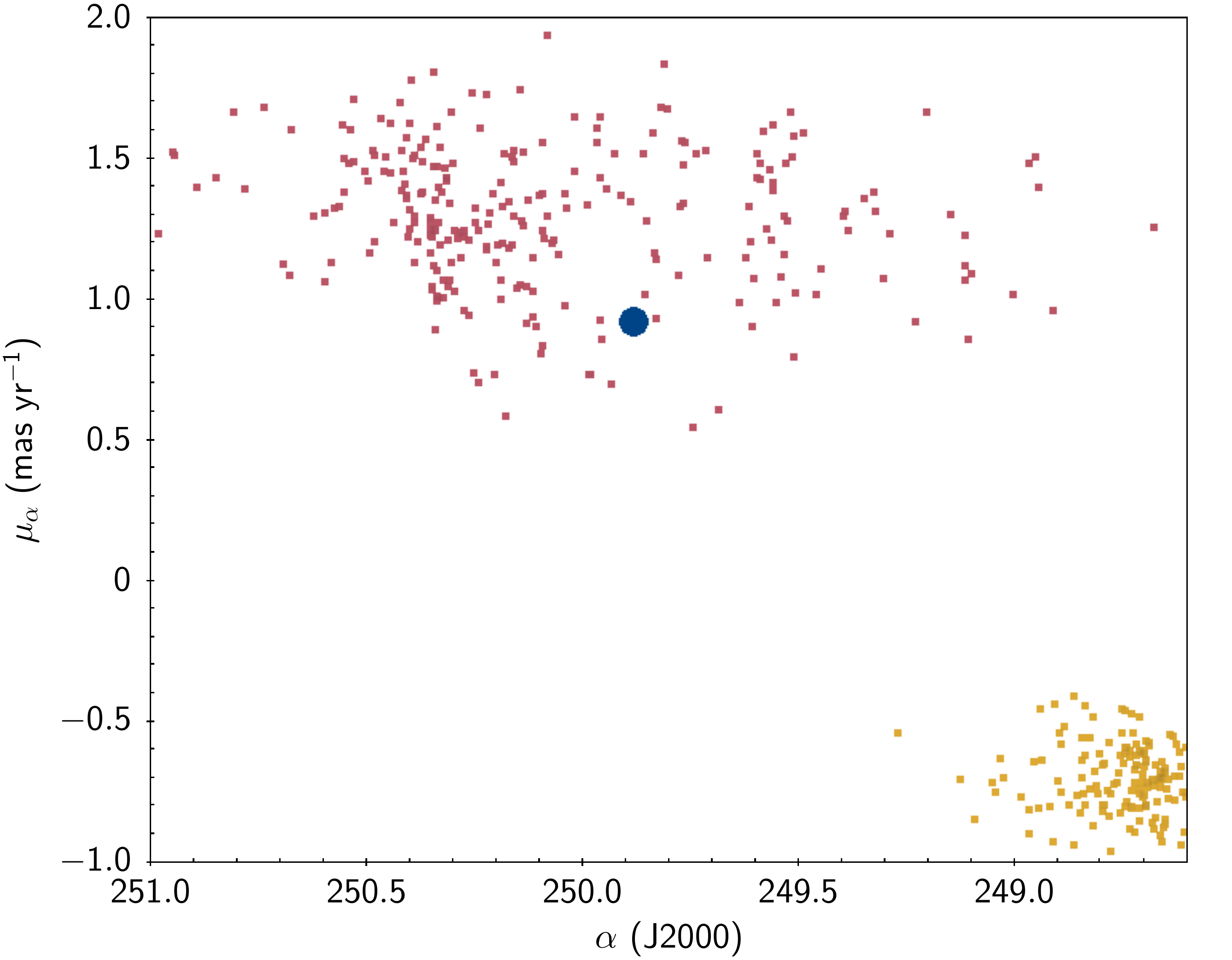}
    \includegraphics[width=0.49\linewidth]{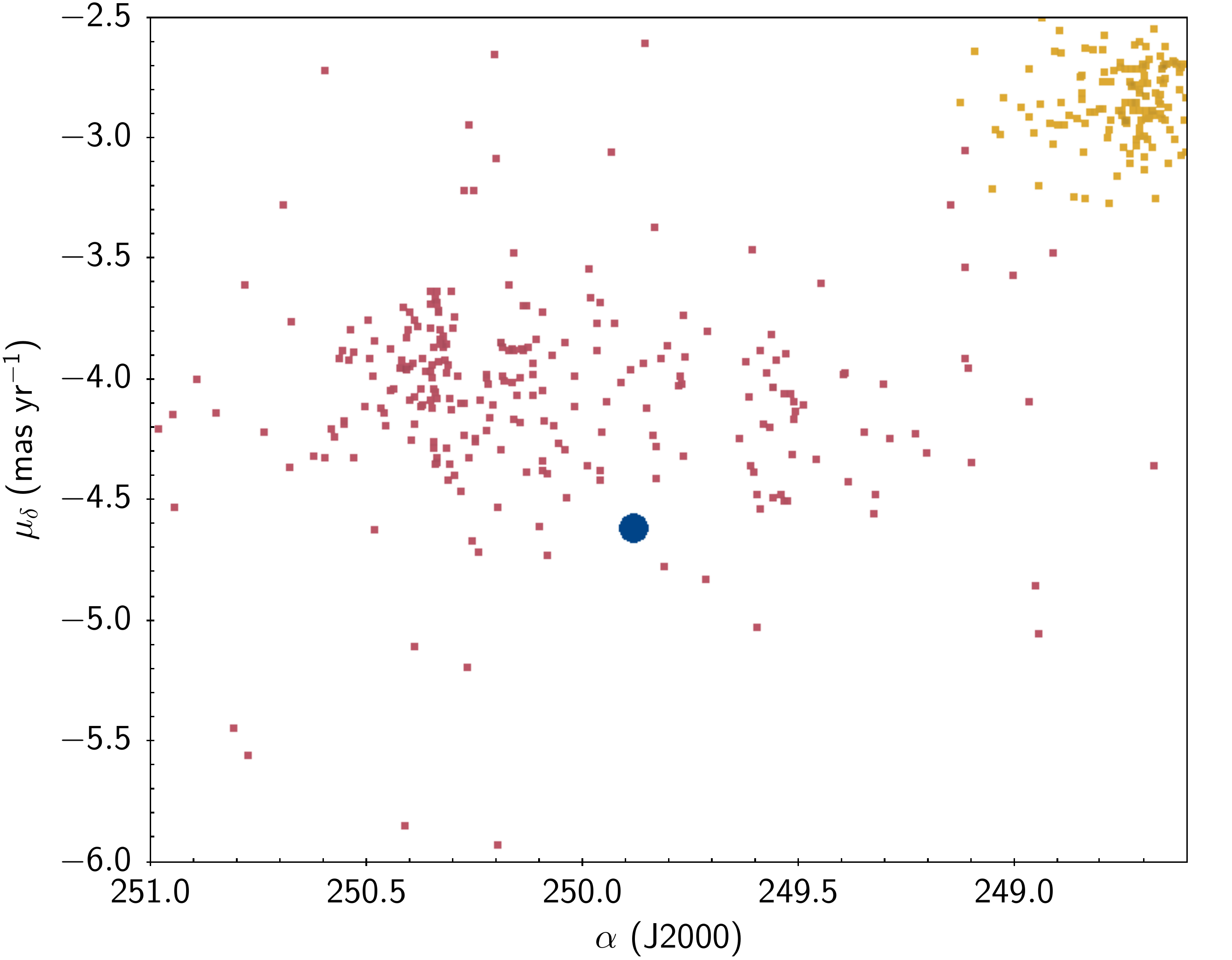}
    \caption{Astrometric and kinematic assessment of membership in NGC 6193 using the catalog from \citet{Kounkel:2020}.}
    \label{fig:membership}
\end{figure*}

At a distance of $\sim$1~kpc, the NGC~6193 cluster is part of the larger Ara OB1a association, which has other clusters also reportedly associated with it. In particular, two other clusters have been recognized as members of this association in the literature: NGC~6167, which is estimated to be 20--30~Myr \citep{baume:2011} but at a significantly greater distance according to {\it Gaia\/} DR2 ($\sim$1400~pc), and NGC~6204, which has a similar distance to NGC~6193 but is estimated to be even older at $\sim$60~Myr \citep{Kounkel:2020}. Accordingly, \citet{baume:2011} suggested a sequential star-formation scenario for Ara OB1 in which NGC~6167 triggered the formation of NGC~6193. 
But with the advent of {\it Gaia\/} DR2 it is now more clear that these clusters' characteristic proper motions are also very different relative to one another, raising doubts as to whether all three clusters are in fact actually related, and in any case presents a clear case for HD~149834 being much more likely to be associated with NGC~6193 (see Fig.~\ref{fig:membership}).  

\begin{figure}[!ht]
    \centering
    \includegraphics[width=\linewidth]{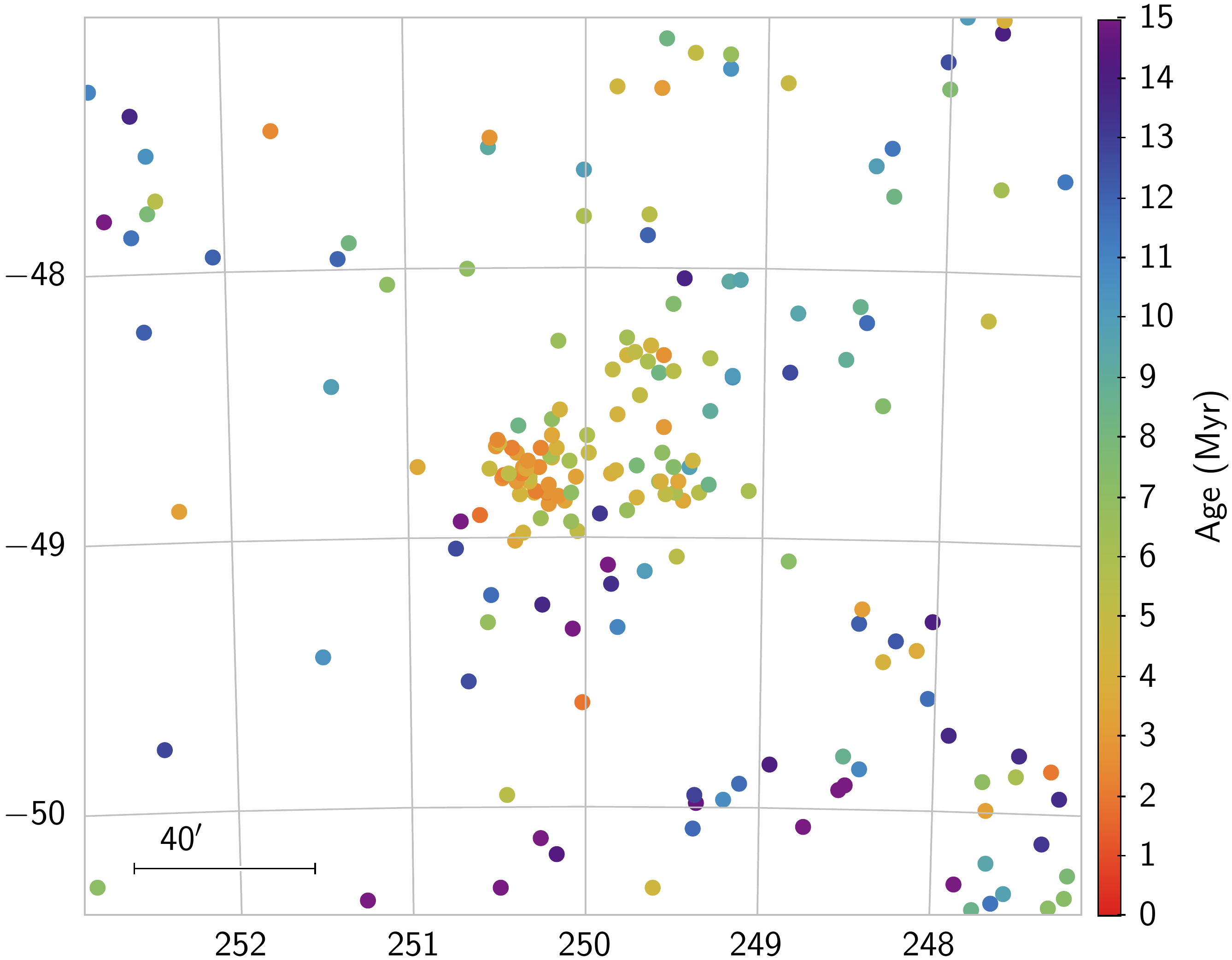}
    \caption{Map of pre-main sequence stars in the direction of NGC 6163, color coded by their estimated age, from McBride et al.\ (submitted).}
    \label{fig:agemap}
\end{figure}

The recent detailed analysis of stellar structures in the local Milky Way by McBride et al.\ (submitted) provides a more granular assessment of individual stellar ages within the region. That analysis finds that while the core of the NGC~6193 cluster has a most likely age of a few Myr, the cluster also appears to possess a ``halo" of somewhat older stars, with ages of 10--15~Myr. 
HD~149834 appears in position and kinematics to be most likely associated with this somewhat older ``halo" of NGC~6193 (see Figs.~\ref{fig:membership}, \ref{fig:agemap}), which could help to resolve the age tension in the CMD noted above (Figure~\ref{fig:cmd}). 

\begin{figure}[!ht]
    \centering
    \includegraphics[width=\linewidth]{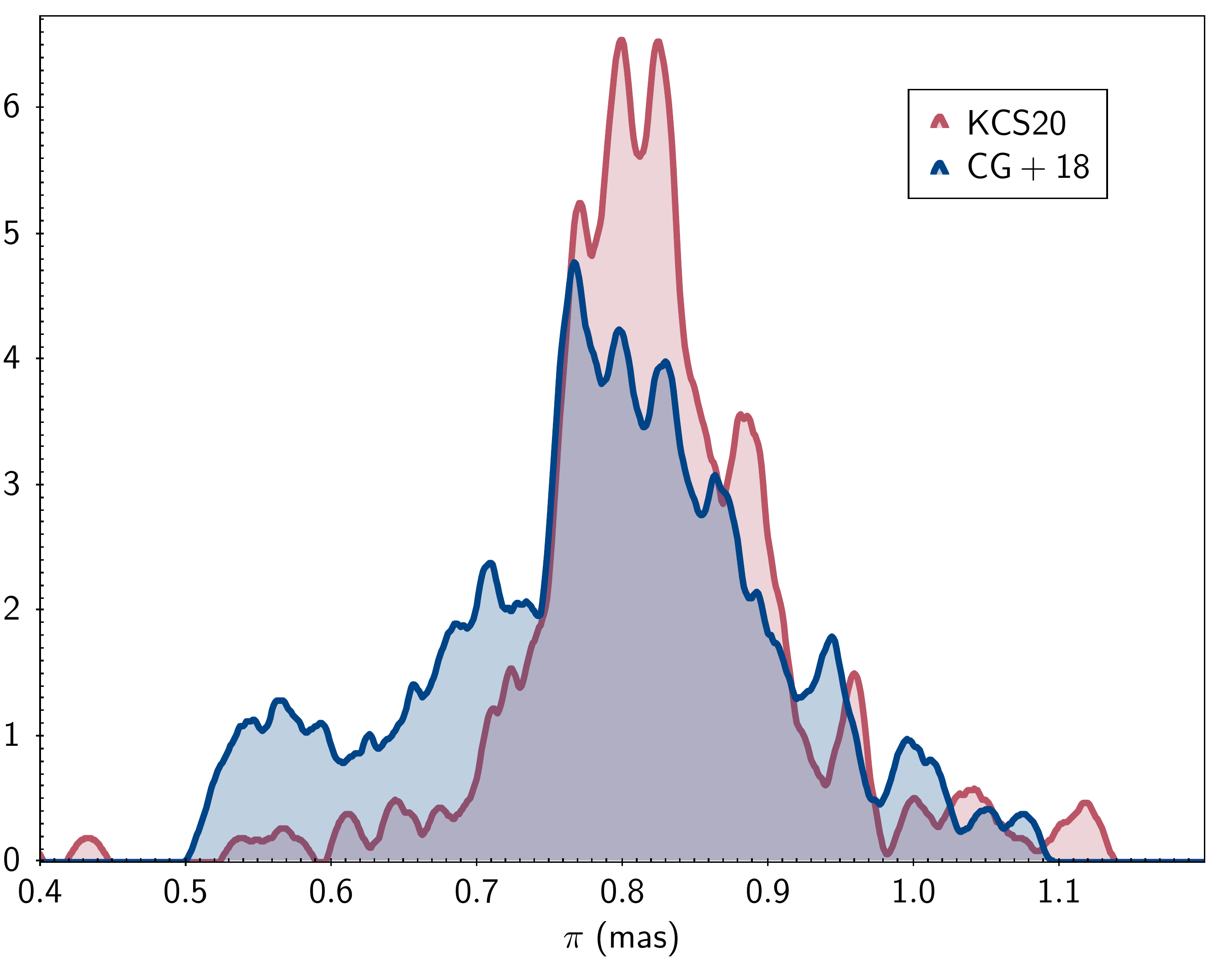}
    \caption{Distribution of parallaxes for NGC 6163 cluster, using the membership from  \citet{Kounkel:2020} and \citet{Cantat-Gaudin:2018}.}
    \label{fig:plx_dist}
\end{figure}

Regarding distance, even with the benefit of {\it Gaia\/} DR2 parallaxes the distance to NGC~6193 is somewhat uncertain due to the small parallax ($\sim$1~mas) and therefore the dependence on the choice of prior. As shown in Figure~\ref{fig:plx_dist}, the 
average {\it Gaia\/} distance to the cluster depends explicitly on which stellar members are used, how the averaging is done, and what systematics are taken into account. This leads to a choice of three distance estimates currently available:
$1055\pm 70 $~pc using the membership from \citet{Kounkel:2020}, 
$1302\pm 108$~pc using the \citet{Kounkel:2020} algorithm on the membership from \citet{Cantat-Gaudin:2018}, or
$1190\pm 145$~pc using the \citet{Cantat-Gaudin:2018} estimate. 

The \citet{Kounkel:2020} analysis gives a roughly symmetric distribution that is quasi-normal and therefore lends itself to a simple $\sigma/\sqrt{N}$ estimate of the uncertainty on the mean based on a Gaussian approximation. 
In that case, the mean parallax for NGC~6193 is $0.835\pm 0.009$~mas. As HD~149834 is a likely member, and as the Gaia parallax measure for the star itself is significantly more uncertain, we adopt this mean cluster parallax as our estimator for the distance to HD~149834.

\section{TESS Light Curve of HD~149834}\label{sec:data} 

\begin{figure*}[!ht]
    \centering
    \includegraphics[width=\linewidth,trim=0 7 0 17,clip]{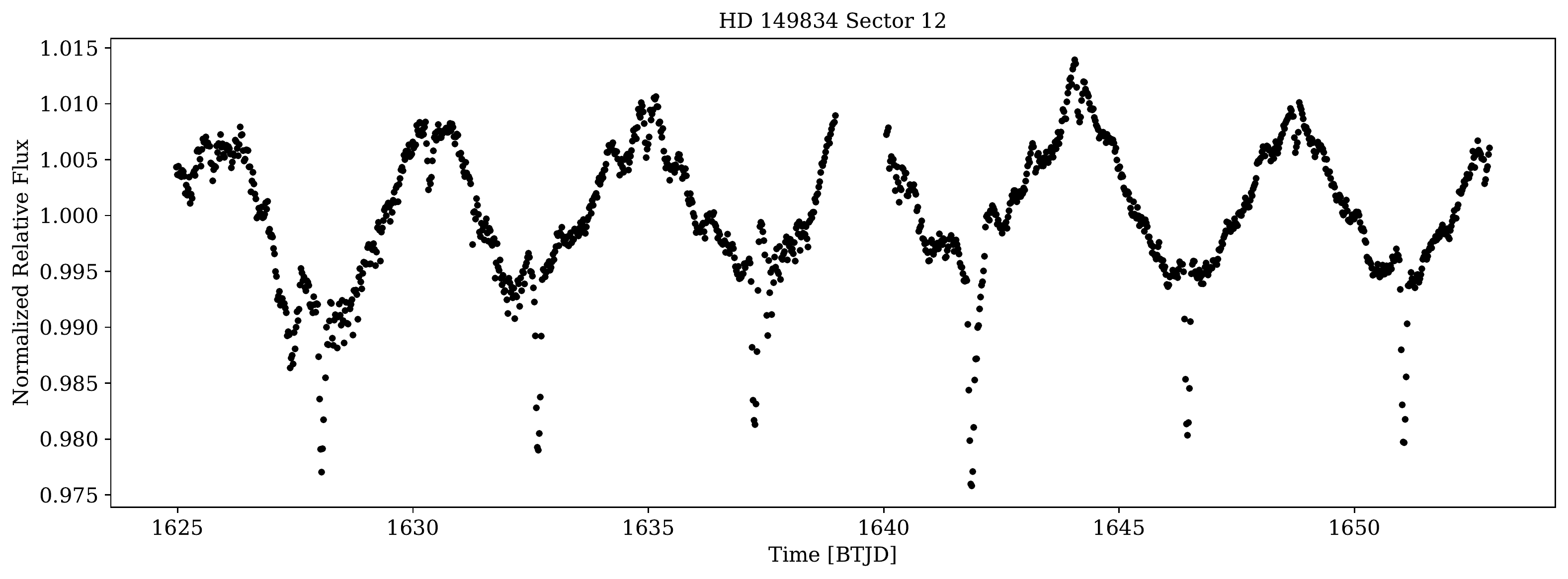}
    \caption{Raw extracted TESS FFI light curve.\\ }
    \label{fig:TESS_lc}
\end{figure*}

\begin{figure}[!ht]
    \centering
    \includegraphics[width=\linewidth,trim=0 7 0 60,clip]{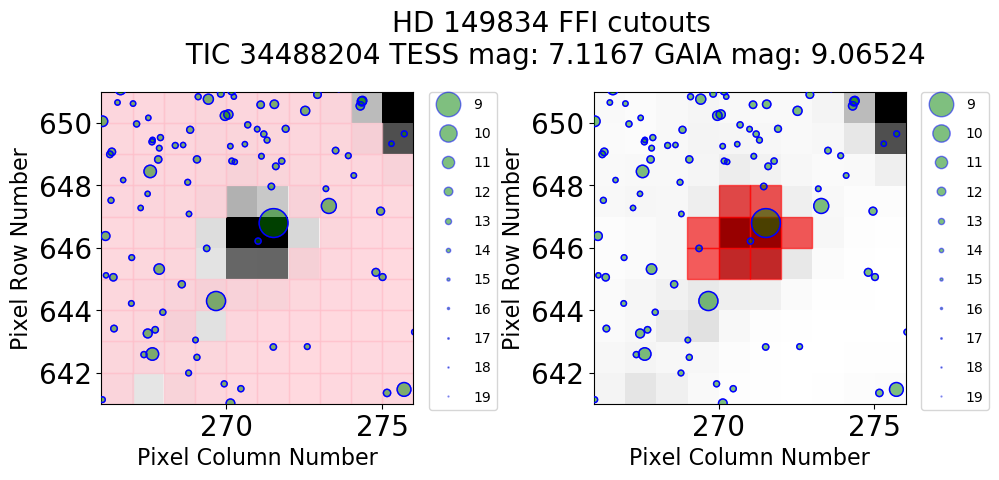}
    \caption{Cutouts of Full Frame Images from TESS Sector 12, centered on HD~149834. Nearby stars from the Gaia catalog are displayed as small circles where the size varies with Gaia magnitudes. In the left panel, the pink color represents the pixel mask to measure the background flux. In the right panel, the red color represents the pixel mask used to extract photometry from the target star.}
    \label{fig:TESS_mask}
\end{figure}

TESS observed HD~149834 in its 30-min full-frame image (FFI) mode over 27~d in Sector~12, showing 
HD~149834 clearly to be an eclipsing binary (EB) with a primary eclipse depth of 1.4\% and a secondary eclipse depth of 0.3\% that could be partial or total (Figure~\ref{fig:TESS_lc}). The $\sim$4.5~d orbital period is consistent with the one from the spectroscopic orbital fit by \cite{Arnal:1988}.
The pixel mask used to extract the light curve from the TESS FFIs is shown in Figure~\ref{fig:TESS_mask}.  


Unfortunately there is no flux contamination estimate provided by the TESS Input Catalog \citep[TIC;][]{StassunTIC:2019} because this star was not included in the TESS Candidate Target List \citep[CTL;][]{StassunTIC:2019}. There is one other relatively bright star in {\it Gaia\/} DR2, a $G=12.8$~mag star that is 38" away. That source is $\sim$3.8~mag fainter and only 1 or 2 pixels away, so its light is surely included in the extracted light curve, contributing $\sim$3\%. All of the other stars that could be contributing light are fainter than $G=17$, but there are enough of them that together they add up to another $\sim$1\%. We conclude that the total dilution of the HD~149834 light curve by extrinsic sources is $\approx$4\%.

The light curve also exhibits out-of-eclipse variations that include a combination of instrumental systematics and true source modulations. It is important to preserve the intrinsic source modulations as they arise from reflection effects and therefore provide important constraints on the physical model solution (see Section~\ref{sec:results}). We therefore opted to detrend the light curve manually, in the following way: Data points were selected by eye near the middle of each ascending and descending branch, and they were fitted with a 5th-order polynomial. This polynomial was divided into the raw fluxes. We then ran a preliminary light curve model (see Section~\ref{subsec:lcfit}) and noticed that the residuals still showed systematics that were a relatively smooth function of time, particularly at the ends of the data set, most likely because the polynomial fit is not able to perform optimally at the ends of the light curve. We therefore fit a spline function to remove just the low-frequency variations in the residuals, and divided this fit into the normalized data from the previous step. This is the final photometry that we will use below in Section~\ref{subsec:lcfit}.




\section{Analysis and Results}\label{sec:results} 




\subsection{Spectral Energy Distribution: Initial Constraints on Stellar Properties}\label{subsec:sed} 

In order to obtain initial estimates of the component effective temperatures ($T_{\rm eff}$) and radii ($R$), we performed a multi-component fit to the combined-light, broadband spectral energy distribution (SED) of the HD~149834 system, starting with the $T_{\rm eff}$ and $R$ estimates for the primary star from the spectroscopic study of \citet{Huang:2006a}. The resulting $T_{\rm eff}$ and $R$ were iteratively updated based on the joint light-curve and radial-velocity model (Section~\ref{subsec:lcfit}) until a final satisfactory SED fit was produced. 

\begin{figure}[!ht]
    \centering
    \includegraphics[width=0.73\linewidth,trim=70 70 90 90,clip,angle=90]{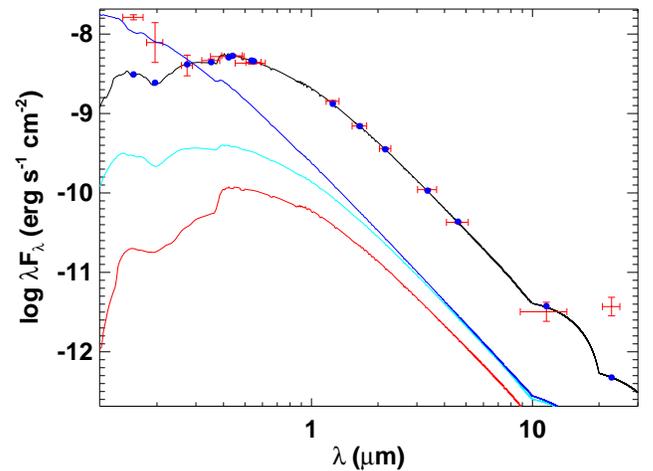}
    \caption{Multi-component fit to the combined-light SED of HD~149834. Observed fluxes are represented as red symbols (horizontal error bars represent filter widths); blue symbols are the corresponding model fluxes. The black curve is a single-source fit using spectroscopic $T_{\rm eff}=21500$~K, [Fe/H]=0, and best-fit $A_V=1.4$. Other curves are attempts to represent the secondary in different ways; the only acceptable choice given the observational constraints on the flux ratio is the red curve with $T_{\rm eff} =$ {13\,000~K} and same $A_V$ as above; see the text.}
    \label{fig:sed}
\end{figure}

As shown in Figure~\ref{fig:sed} (black curve), the SED from 0.3~$\mu$m to 10~$\mu$m can be very well fit by a single component with $T_{\rm eff} = 21467 \pm 246$~K, $\log g = 3.95 \pm 0.15$, and 
[Fe/H]$\,\approx\,$0
as reported from a previous spectroscopic analysis in the literature {\citep{Huang:2006b}}, with a best-fit extinction of $A_V = 1.4 \pm 0.1$ {\citep[consistent with other determinations of the reddening of this cluster; e.g.,][]{Rangwal:2017}}. 

Interestingly, there is an apparent excess in the UV. In principle this could be due to activity on the low-mass companion star; young low-mass stars have ubiquitous UV excess even after they stop accreting. Alternatively, the UV excess could suggest an additional hot source in the system. However, adding a second component cannot reproduce that excess if the same $A_V$ is applied. For example, {adopting the secondary star's radius determined from the light curve analysis (Sec.~\ref{subsec:lcfit}), we can reproduce the UV excess with a secondary $T_{\rm eff} = 25\,000$~K} (the dark blue curve in Fig.~\ref{fig:sed}), however this requires zero extinction. Reddening it by the same $A_V$ as above yields the light blue curve, which predicts gives a flux ratio in the TESS band of {6.9\%}, much too large for the $\sim${2.5\%} flux ratio that is required by the light curve analysis (Section~\ref{subsec:lcfit}). 

Therefore the UV excess is not easily explained, and we note also evidence for an excess at 20~$\mu$m. 
We are able to reproduce the $\sim${2.5\%} flux ratio in the TESS band with the same $A_V$ applied if we instead adopt $T_{\rm eff} = ${13\,000~K} (red curve in Fig.~\ref{fig:sed}). 
Such a faint companion would require extraordinarily strong intrinsic UV emission to produce the observed UV excess, some 5 orders of magnitude over photospheric (see Fig.~\ref{fig:sed}). Thus the origin of the UV emission, if real, remains unclear (but see Sec.~\ref{subsec:chandra} regarding observed X-ray emission in the system). 

Most importantly, the SED analysis provides robust initial constraints on the properties of the stars that we can utilize in the full solution below (Sec.~\ref{subsec:lcfit}). 
Adopting the parallax estimated in Sec.~\ref{sec:hd149834} 
and correcting for the 0.029~mas systematic offset of the {\it Gaia\/} DR2 parallaxes reported by the {\it Gaia\/} mission, we obtain via the bolometric flux from the SED fit an estimate for the primary star radius: 
$R_1 = 4.96 \pm 0.13$~R$_\odot$. 
This radius, together with the spectroscopic $\log g$, then gives an estimate for the primary mass of $M_1 = 8.1 \pm 1.3$~M$_\odot$. 

\subsection{TESS Light Curve Analysis: Detailed Determination of Stellar Properties}
\label{subsec:lcfit}

We used the detrended TESS photometry described in Section~\ref{sec:data}, along with its internal
uncertainties, to perform a light curve analysis of \hd\ using the {\tt eb} code of
\cite{Irwin:2011}, which is based on the Nelson-Davis-Etzel binary model \citep{Etzel:1981, Popper:1981}
underlying the popular EBOP program by those authors. This model is adequate for well-detached systems
such as \hd, having nearly spherical stars. The main adjustable parameters we considered are the orbit
period ($P$), a reference epoch of primary eclipse ($T_0$, which in this code is strictly the time of
inferior conjunction), the central surface brightness ratio in the TESS bandpass ($J \equiv J_2/J_1$),
the brightness level at the first quadrature ($m_0$), the sum of the relative radii normalized by the
semimajor axis ($r_1+r_2$), the cosine of the inclination angle ($\cos i$), and the eccentricity
parameters $e \cos\omega$ and $e \sin\omega$, with $e$ being the eccentricity and $\omega$ the
longitude of periastron. The radius ratio ($k \equiv r_2/r_1$), normally also an adjusted parameter in
light curve solutions, was instead derived at each iteration of our analysis based on other
information, as we explain below. In order to account for the flux contamination in the TESS aperture
described in Sec.~\ref{sec:data}, we also solved for the third light parameter $\ell_3$, defined
such that $\ell_1 + \ell_2 + \ell_3 = 1$, in which $\ell_1$ and $\ell_2$ for this
normalization are taken to be the light at first quadrature.

A linear limb-darkening law was adopted for this work, with coefficients {for solar metallicity} taken from the tabulation by
{\cite{Claret:2017} for the TESS band, interpolated to the temperatures and surface gravities of the stars based on our final analysis.} The same source was used for the gravity darkening coefficients. Due to the large difference
in temperature between the components indicated earlier, there is a strong reflection effect in the
light curve caused by the energy from the primary being re-radiated toward the observer by the cooler secondary star, with
a strength controlled by the reflection coefficient of the secondary, $\eta_2$ \citep{Etzel:1981}. We
therefore solved for this additional parameter, while leaving the corresponding coefficient for the
primary set to zero, as its impact is negligible. We accounted for the finite time of integration of
the TESS photometry by oversampling the model light curve and then integrating over the 30-minute
duration of each cadence prior to the comparison with the observations \citep[see][]{Gilliland:2010,
Kipping:2010}.

In addition to the photometry, we incorporated the radial velocities of \cite{Arnal:1988} directly into
the analysis together with their reported uncertainties, solving for the velocity semiamplitude of the
primary star ($K_1$) and the center-of-mass velocity of the system ($\gamma$). {The use of these older
data extends the time base significantly to nearly 35 years, thereby improving the period determination.}

The solution was carried out within a Markov chain Monte Carlo (MCMC) framework using the {\tt
emcee\/}
code of \cite{Foreman-Mackey:2013}, with 100
walkers of length 15,000 each, after discarding the burn-in. We adopted uniform priors over suitable
ranges for most parameters mentioned above, and a log-uniform prior for $\eta_2$. The prior for
$\ell_3$ was assumed to be Gaussian, with a mean value of 4\% as derived in Section~\ref{sec:data}, and
a standard deviation of 1\%. Convergence of the chains was checked visually, requiring also a
Gelman-Rubin statistic of 1.05 or smaller for each parameter \citep{Gelman:1992}.  The relative
weighting between the photometry and the radial velocities was handled by including additional
adjustable parameters $f_{\rm phot}$ and $f_{\rm RV}$ to rescale the observational errors.  These
multiplicative scale factors were solved for self-consistently and simultaneously with the other
orbital quantities \citep[see][]{Gregory:2005}, adopting log-uniform priors for both.

As we only have radial velocities for the primary star, an estimate of the mass $M_1$ of that component
is required before we can derive the absolute the mass and radius of the secondary. The spectroscopic
study of \cite{Huang:2006b} reported an estimate of the surface gravity for the primary of $\log g_1 =
3.955 \pm 0.023$, {although they considered 0.15 dex to be a more realistic estimate of the uncertainty. We adopt this more conservative error here. The $\log g$ value} together with the radius estimate $R_1$ for that star from our SED fit
above yields the necessary information to infer $M_1$. In order to propagate the uncertainties more
accurately to the radius and mass of the secondary, we chose to add $\log g_1$ and $R_1$ as adjustable
parameters in our solution, constrained by Gaussian priors given by the measured values and
corresponding errors of those two quantities. At each iteration we then solved for the mass and radius
of the secondary, which immediately provides the radius ratio $k \equiv r_2/r_1 = R_2/R_1$ needed to
model the light curve.

\begin{figure*}
    \centering
    \includegraphics[width=0.59\linewidth,angle=270,trim=0 0 200 0,clip]{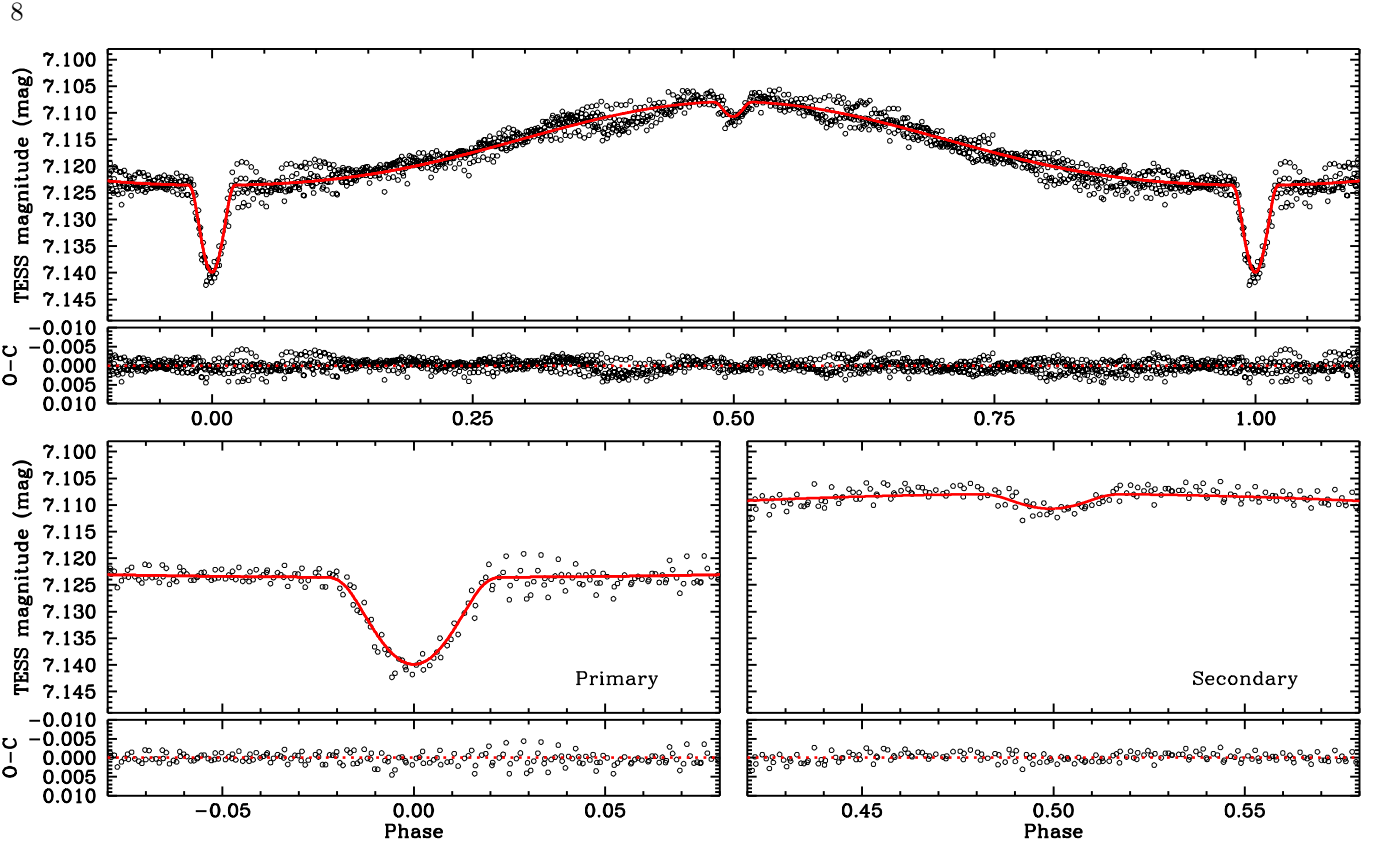}
    \caption{Final detrended TESS photometry with our model fit shown in red.}
    \label{fig:lcfit}
\end{figure*} 

The results of our analysis for \hd\ are presented in Table~\ref{tab:mcmc}, in which we report the mode
of the posterior distribution for each parameter. Posterior distributions of the derived quantities
listed in the bottom section of the table were constructed directly from the MCMC chains of the
adjustable parameters involved. Our adopted lightcurve model along with the TESS observations are seen
in Figure~\ref{fig:lcfit}, with residuals shown at the bottom of each panel.

\setlength{\tabcolsep}{2pt}
\begin{deluxetable}{lcc}
\tablecaption{Results From our MCMC Analysis of \hd \label{tab:mcmc}}
\tablehead{ \colhead{~~~~~~~~~Parameter~~~~~~~~~} & \colhead{Value} & \colhead{Prior} }
\startdata
$P$ (days)                       &  $4.595701^{+0.000035}_{-0.000038}$   & [4, 5] \\ [1ex]
$T_0$ (HJD$-$2,400,000)          &  $58641.8547^{+0.0010}_{-0.0011}$     & [58641, 58642] \\ [1ex]
$J$                              &  $0.21^{+0.23}_{-0.11}$               & [0.0, 1.0] \\ [1ex]     
$r_1+r_2$                        &  $0.249^{+0.014}_{-0.016}$            & [0.01, 0.60] \\ [1ex]
$\cos i$                         &  $0.224^{+0.015}_{-0.017}$            & [0, 1] \\ [1ex]
$e \cos\omega$                   &  $-0.0009^{+0.0014}_{-0.0017}$        & [$-$1, 1] \\ [1ex]
$e \sin\omega$                   &  $-0.037^{+0.012}_{-0.014}$           & [$-$1, 1] \\ [1ex]
$m_0$ (mag)                      &  $7.12210^{+0.00014}_{-0.00020}$      & [7.0, 7.5] \\ [1ex]
$\eta_2$                         &  $0.00777^{+0.00012}_{-0.00011}$      & [$-15$, $-1$] \\ [1ex]
$\ell_3$                         &  $0.0399^{+0.0103}_{-0.0095}$         & $G(0.04, 0.01)$ \\ [1ex]
$\gamma$ (\kms)                  &  $-34.7^{+1.7}_{-1.8}$                & [$-50$, $-20$] \\ [1ex]
$K_1$ (\kms)                     &  $22.5^{+1.9}_{-2.2}$                 & [10, 50] \\ [1ex]
$f_{\rm phot}$                   &  $8.54^{+0.19}_{-0.15}$               & [$-5$, 5] \\ [1ex]
$f_{\rm RV}$                     &  $2.17^{+0.57}_{-0.26}$               & [$-5$, 5] \\ [1ex]
$R_1$ ($R_{\sun}$)               &  $4.96^{+0.12}_{-0.14}$               & $G(4.96, 0.13)$ \\ [1ex]
$\log g_1$ (cgs)                 &  $4.040^{+0.034}_{-0.029}$            & $G(3.955, 0.15)$ \\ [1ex]
\noalign{\hrule} \\ [-2.5ex]
\multicolumn{3}{c}{Derived quantities} \\ [1ex]
\noalign{\hrule} \\ [-2.5ex]
$r_1$                            &  $0.1940^{+0.0035}_{-0.0054}$         & \nodata \\ [1ex]
$r_2$                            &  $0.0468^{+0.0231}_{-0.0065}$         & \nodata \\ [1ex]
$k \equiv r_2/r_1$               &  $0.235^{+0.129}_{-0.026}$            & \nodata \\ [1ex]
$i$ (degrees)                    &  $77.07^{+1.02}_{-0.86}$              & \nodata \\ [1ex]
$e$                              &  $0.037^{+0.012}_{-0.014}$            & \nodata \\ [1ex]
$\omega$ (degrees)               &  $91.0^{+3.7}_{-1.8}$                 & \nodata \\ [1ex]
$K_2$ (\kms)                     &  $251.6^{+11.0}_{-7.4}$               & \nodata \\ [1ex]
$M_1$ ($M_{\sun}$)               &  $9.71^{+1.22}_{-0.81}$               & \nodata \\ [1ex]
$M_2$ ($M_{\sun}$)               &  $0.858^{+0.123}_{-0.096}$               & \nodata \\ [1ex]
$q \equiv M_2/M_1$               &  $0.0874^{+0.0093}_{-0.0079}$         & \nodata \\ [1ex]
$R_2$ ($R_{\sun}$)               &  $1.16^{+0.64}_{-0.13}$               & \nodata \\ [1ex]
$\log g_2$ (cgs)                 &  $4.11^{+0.23}_{-0.24}$               & \nodata \\ [1ex]
$a$ ($R_{\sun}$)                 &  $25.57^{+0.97}_{-0.76}$              & \nodata \\ [1ex]
\enddata
\tablecomments{The values listed correspond to the mode of the respective posterior distributions,
and the uncertainties represent the 68.3\% credible intervals. All priors are uniform over the
specified ranges, except those for $\eta_2$, $f_{\rm phot}$, and $f_{\rm RV}$, which are log-uniform,
and the priors for $R_1$, $\log g_1$, and $\ell_3$, which are Gaussian with mean and standard
deviations as indicated by the notation $G({\rm mean}, \sigma)$.} 

\end{deluxetable} 
\setlength{\tabcolsep}{6pt}

\begin{figure}[!ht]
    \includegraphics[width=\linewidth,trim=0 0 125 440,clip]{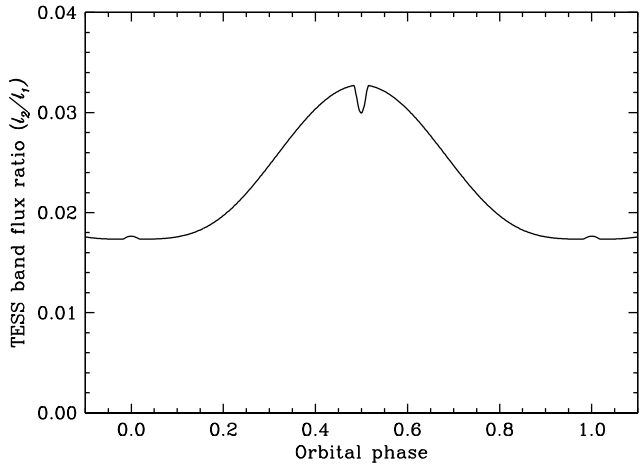}
    \caption{Light ratio between the secondary and primary components of \hd\ as a function of orbital
phase, showing the significant changes due to the reflection effect.}
    \label{fig:lcphase}
\end{figure}

\begin{figure}[!ht]
    \includegraphics[width=\linewidth,trim=0 0 120 380,clip]{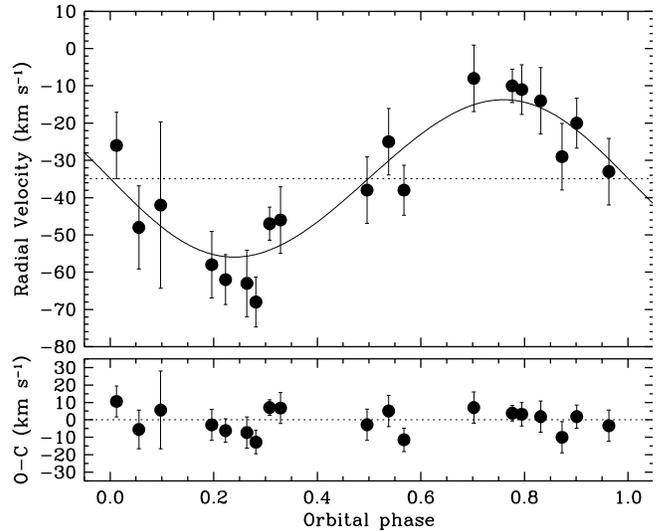}
    \caption{Radial velocity measurements of \cite{Arnal:1988} along with our model fit. Residuals
are displayed at the bottom.}
    \label{fig:RVs}
\end{figure}

Due to the reflection effect the light ratio between the components of \hd\ changes by a
factor of about {1.9} as a function of orbital phase. This is illustrated in
Figure~\ref{fig:lcphase}. 
Our model radial-velocity curve with the observations of
\cite{Arnal:1988} is shown in Figure~\ref{fig:RVs}, indicating a center-of-mass velocity consistent with membership in NGC~6193, as mentioned earlier.

\subsection{HR Diagram: Evolutionary Analysis}\label{subsec:hrd} 

\begin{figure}[!ht]
    \centering
    \includegraphics[width=\linewidth,trim=100 410 80 100,clip]{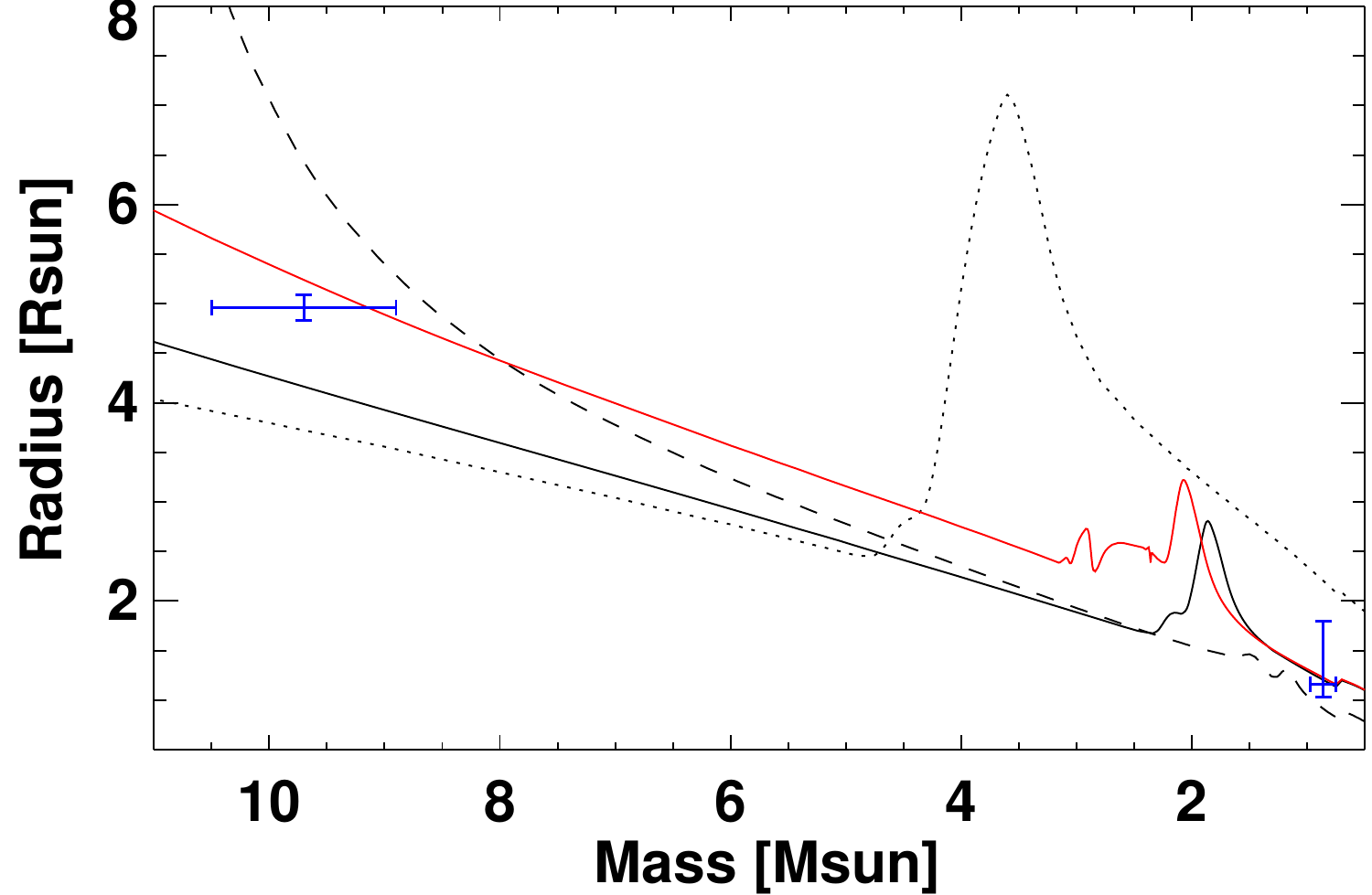}
    \includegraphics[width=\linewidth,trim=100 80 80 420,clip]{hd149834_hrd_fin.pdf}
    \caption{Mass-radius (top) and $T_{\rm eff}$-radius (bottom) representations of the HD~149834 primary and secondary (blue error bars). Black curves are PARSEC isochrones at solar metallicity for ages of 1~Myr (dotted), 5~Myr (solid), and 20~Myr (dashed). Red curves are the same but for [M/H] = $+$0.7 at the nominal cluster age of 5~Myr.}
    \label{fig:hrd}
\end{figure}

Figure~\ref{fig:hrd} depicts the two components of the HD~149834 system in the mass-radius and $T_{\rm eff}$-radius planes, compared to the PARSEC model isochrones. 
In the mass-radius plane, the secondary appears roughly as expected for the nominal cluster age of 5~Myr, whereas the primary appears too large for that age; it appears slightly evolved with an inferred age of $\sim$20~Myr. 

In the $T_{\rm eff}$-radius plane, both components can be reasonably well fit by the 20~Myr isochrone {(note the very large uncertainty on the secondary $T_{\rm eff}$ due to the large uncertainty on $J$ from the light curve analysis)}. However, this is misleading because the secondary's measured mass of $\sim${0.86}~M$_\odot$ should place it at a much cooler $T_{\rm eff}$. 

It is possible to explain both stars in the mass-radius diagram at about 5~Myr with a super-solar metallicity; however the required metallicity is quite high at +0.7 (red curve). {Unfortunately,
there is no reliable literature estimate of the metallicity for \hd.}
Nonetheless, given all of the available evidence, we prefer this solution as both stars are then most consistent with the nominal age of the NGC~6193 cluster, and the anomalously high $T_{\rm eff} = 13\,000^{+7500}_{-4000}$~K for the secondary attributed to its extremely high irradiation from the primary.

\subsection{Chandra X-ray Light Curve: Activity}\label{subsec:chandra} 

\begin{figure}[!ht]
    \centering
    \includegraphics[width=\linewidth,trim=25 10 15 15,clip]{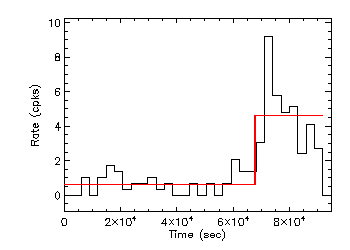}
    \caption{Chandra X-ray light curve (black) and Bayesian block (red) for HD~149834 (S.~Wolk, private communication).}
    \label{fig:chandra}
\end{figure}

\citet{Wolk:2008} observed HD~149834 (Source 174 in their study) using Chandra for $\sim 25$ hours, starting at UT 2:37 on 2004 October 25. They identified HD~149834 as a flare source using Bayesian blocks \citep{Scargle:1998} with variability detected at 95\% significance; its light curve is plotted in Figure~\ref{fig:chandra}. Based on the ephemeris we determine in Section~\ref{subsec:lcfit} (see Table~\ref{tab:mcmc}), the {\it Chandra} observations span orbital phases {0.57--0.80, with a {formal} uncertainty of less than 0.01}. {However, the possibility of multiple period aliases makes the uncertainty likely larger.}

{Taken at face value, the ephemeris suggests that} the observed flare {occurs near phase 0.75}, when HD~149834\,B is accelerating towards us. {In any case,} along with the UV excess seen in the SED (Sec.~\ref{subsec:sed} and Fig.~\ref{fig:sed}), this flare may be a signature of activity from HD~149834. An actual detection of the secondary eclipse in the X-ray light curve would have implicated HD~149834\,B as the source.

\subsection{TESS Residuals: Pulsations and Rotation}\label{subsec:pulsations} 

\begin{figure}[!ht]
    \centering
    \includegraphics[width=\linewidth,trim=15 10 10 10,clip]{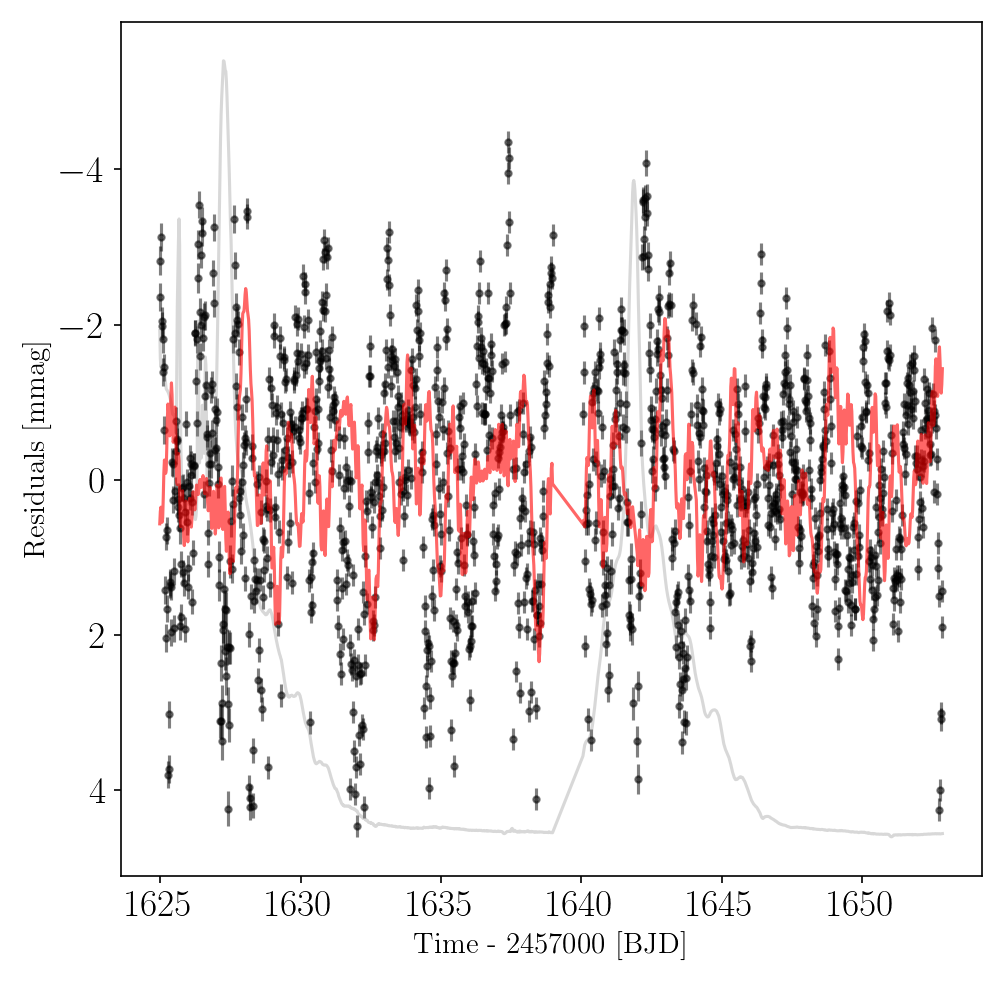}
    \caption{Residuals (black) and pulsation model (red) including all significant frequencies listed in Table~\ref{tab:frequencies}. Re-scaled background lightcurve in grey. }
    \label{fig:residuals_lc}
\end{figure}

Following the subtraction of the best fit model determined in Section~\ref{subsec:lcfit}, we investigate 
the residuals to determine if any remaining signal is present. By comparing the residuals (black) with 
the background (grey) in Fig.~\ref{fig:residuals_lc}, we see that the variation induced by the Earth-shine
events is still present, despite the detrending. This event occurs twice, separated by approximately 16 
days. 

We subjected the residuals to an iterative pre-whitening analysis using the {\sc pythia} python package
wherein we calculate a Lomb-Scargle 
periodogram of the residuals, identify the frequency with the highest amplitude, and fit a sinusoid 
to the data to determine the optimal frequency, amplitude, and phase of that signal. The optimized
signal is then subtracted, and the signal-to-noise ratio (S/N) of signal is computed by dividing the
amplitude of the optimized sinusoid by the average noise level of the {periodogram of the residuals 
after subtraction of the identified signal within a window of 3 d$^{-1}$. This process is repeated 
until an extracted peak is found to have S/N $<$ 3.5 At the end of this process, we simultaneously 
fit all frequencies, amplitudes, and phases to the original data, and re-calculate the S/N according 
to the new residuals. We retain all independent frequencies with S/N $>$ 4 \citep{Breger:1993}. Following
\citet{Breger:1999} and considering the average uncertainty of the extracted frequencies, we retain any 
combination frequency with S/N $>$ 3.3.} 

\begin{figure}[!ht]
    \centering
    \includegraphics[width=\linewidth,trim=10 10 5 5,clip]{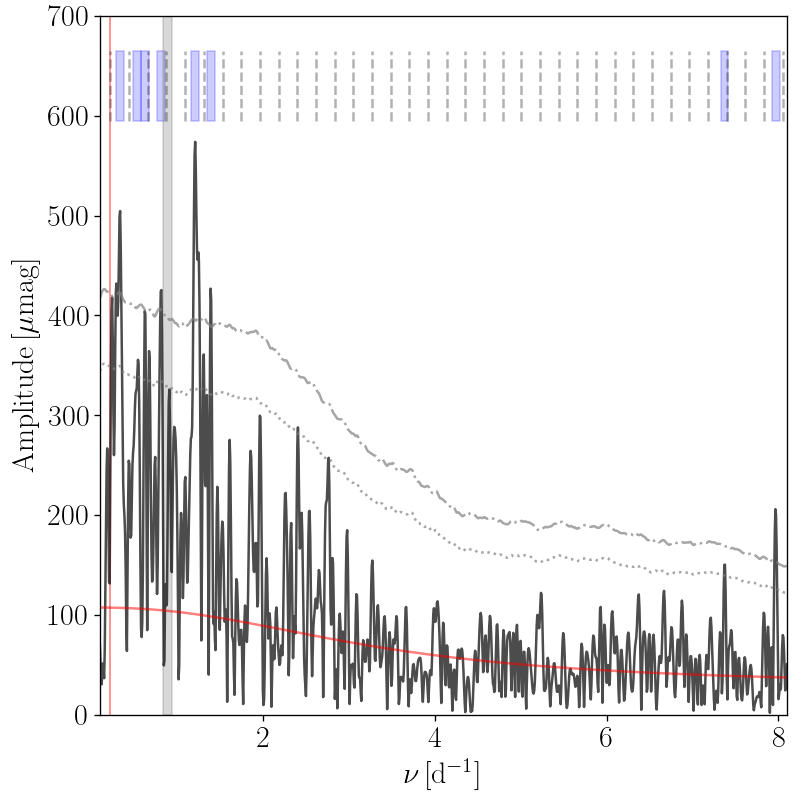}
    \caption{Lomb-Scargle periodogram of the residuals after subtraction of the binary model. Grey dash-dotted and grey dotted lines indicate the a factor 3.5 and 3 times the noise level, respectively. Grey vertical region denotes the rotational period calculated using the values in Table~\ref{tab:mcmc}. Vertical blue regions denote the extracted frequencies and their uncertainties, and vertical dashed grey lines denote orbital harmonics. Red profile is the red-noise profile fit to the periodogram of the residuals after iterative prewhitening. }
    \label{fig:residuals_ls}
\end{figure}

Following \citet{Degroote:2009}, after extraction we filter for close frequencies defined as $\delta f 
\le 2.5 f_{\rm R}$, and keep the frequency which has the higher amplitude. Here, the Rayleigh frequency, 
$f_{\rm R}\approx0.036$~d$^{-1}$ is the inverse of the time-base of the TESS lightcurve. We identify {8} 
frequencies, listed in Table~\ref{tab:frequencies}, {two} of which are combination frequencies or orbital
harmonics, within the frequency uncertainties. The full Lomb-Scargle periodogram of the binary subtracted 
residuals is shown in Fig.~\ref{fig:residuals_ls}, where {4 and 3.3} times the average noise 
level are depicted by the dashed-dotted {and dotted grey lines, respectively}. {Furthermore, 
we identify $f_{7}$ as matching both the 34th orbital harmonic.}
The residuals after binary subtraction and optimised sinusoid model comprised of the frequencies in 
Table~\ref{tab:frequencies} are shown in Fig.~\ref{fig:residuals_lc} in black and red, respectively. 

{Additionally, we note that the mass of the primary is within the range of stars that are expected
to exhibit internal gravity waves (IGWs) excited at the boundary of the convective core. Both 2D and 3D
hydrodynamical simulations demonstrate that IGWs observationally manifest as a low-frequency excess and/or 
as resonantly excited g modes in photometry \citep{Rogers:2013, Bowman:2019b, Lecoanet:2019, Edelmann:2019, 
Horst:2020}. Given the observed low-frequency excess in the periodogram of the residuals of HD~149834, we 
fit periodogram according to \citet{Blomme:2011,Bowman:2019b}, with:
\begin{equation}
    \label{eqn:bkg}
    \alpha_{\nu} = \frac{\alpha_0}{1 + \left( \nu / \nu_{\rm char}\right)^{\gamma}}+W,
\end{equation}
to determine the parameters of the background profile. Here, $\alpha_0$ and $W$ are the amplitude
at $\nu =0$ and the white noise, respectively. $\gamma$ controls the slope of the profile, and $\nu_{\rm char}$
is the characteristic frequency, at which the profile has half of its initial value. We find that the 
low-frequency excess is best described with $W=28.6~\mu$mag, $\alpha_0=78.8~\mu$mag, $\gamma=2.34$,
and $\nu_{\rm char}=3.32$~d$^{-1}$. The best fit profile is plotted in red in Fig.~\ref{fig:residuals_ls}.}

\begin{table}[!ht]
    \centering
    \begin{tabular}{lcccc}
    \hline
    & Frequency & Amplitude & SNR & Comment \\
    & d$^{-1}$  &   mmag    &     &         \\
    \hline
    $f_1$    & $0.329\pm0.0171$ & $0.56\pm0.00$ & 5.3 & \\
    $f_2$    & $0.527\pm0.0216$ & $0.44\pm0.00$ & 4.3 & \\
    $f_3$    & $0.618\pm0.0282$ & $0.34\pm0.00$ & 3.3 & $f_6-f_4$\\
    $f_4$    & $0.806\pm0.0235$ & $0.41\pm0.00$ & 4.0 & \\
    $f_5$    & $1.201\pm0.0158$ & $0.60\pm0.00$ & 6.2 & \\
    $f_6$    & $1.390\pm0.0276$ & $0.35\pm0.00$ & 4.0 & \\
    $f_7$    & $7.373\pm0.0662$ & $0.14\pm0.00$ & 3.5 & $34f_{\rm orb}$\\
    $f_8$    & $7.969\pm0.0459$ & $0.21\pm0.00$ & 5.2 & \\

    \hline
    \end{tabular}
    \caption{Extracted and filtered frequencies from iterative pre-whitening analysis.}
    \label{tab:frequencies}
\end{table}

\section{Discussion}\label{sec:discussion} 

\subsection{Cluster membership and age considerations}

The kinematics, distance, and age of HD~149834 are consistent with it being a member of 
NGC~6193. The mass of the primary makes it one of the most massive members of this cluster.
HD~149834\,A is at the mass limit where a star is predicted to produce a supernova at the 
end of its life, depending on the rotational and internal chemical mixing history of the
star. Notably, HD~149834 is not the most massive system in NGC~6193. There are two massive 
multiple systems, HD~150135 \citep[a binary;][]{Sota:2014} and HD~150136 \citep[a triple;][]
{Mahy:2012,Sana:2012,Mahy:2018}, both of which are confirmed luminous X-ray sources
\citep{Skinner:2005}. Using a combination of radial velocities \citep{Arnal:1988,Mahy:2012,
Sana:2013,Cox:2017}, photometry, and long time-base interferometry \citep{Sanchez-Bermudez:2013,
LeBouqin:2017}, \citet{Mahy:2018} found HD~150136 to be comprised of a $\sim42.8+29.5$~M$_{\odot}$
close inner binary and $\sim15.5$~M$_{\odot}$ wide tertiary. Furthermore, they determine a common
age of the components to be 0--2~Myr, which is about half the nominal age estimate of the cluster. 

As mentioned in Section~\ref{subsec:hrd}, there is some tension between the metallicity and age
of HD~149834, which may be explained either by the anomalously hot secondary or by the system
being a member of the cluster halo, which is estimated to be slightly older than the cluster core.
This, however, puts more contention between the age estimates of HD~150136 and HD~149834. Alternatively,
such an age gradient between halo and core members of the same cluster has been observed in other
young clusters and is evidence for outside-in star formation scenarios \citep{Getman:2014}.

\subsection{Coherent and stochastic oscillations in HD~149834~A}
Of the {8} extracted signals, {6} are independent and {2} are low-order 
linear combination frequencies {or harmonics}. The {extracted} frequencies 
occur in two distinct frequency regimes. {The single independent high frequency 
we extracted is consistent with the expected frequency range for p mode pulsations
in $\beta$~Cep stars. Those lower frequency peaks we extracted occur in a frequency
regime that is consistent with rotational modulation, low-frequency instrumental
noise, low-order heat-driven p  and g modes, as well as IGWs and pulsation modes 
resonantly excited by IGWs.} The {9.7}~M$_{\odot}$ primary is situated in 
the mass range where both g  and p mode pulsations are theoretically expected 
to be excited \citep{Walczak:2015,Moravveji:2016,Szewczuk:2017}. 

{Furthermore, 
the values we determined for $\gamma$ and $\nu_{\rm char}$ are consistent with 
the values expected for IGWs in massive stars as determined from simulations 
\citep{Edelmann:2019,Horst:2020} as well as a large sample of OB type stars 
observed by Kepler, CoRoT, and TESS \citep{Bowman:2020a}. While subsurface 
convection has been proposed to generate similar signals \citep{Cantiello:2009,
Lecoanet:2019}, it is expected to produce low-frequency profiles with significantly 
different values of $\gamma$, i.e., $\gamma\geq3.5$ \citep{Couston:2018}, than 
those expected to be produced by IGWs, i.e., $1 \le \gamma \le 3.5$ \citep{Edelmann:2019}.
To this end, we find that the low-frequency power excess observed in the TESS 
data of HD~149834 to be consistent with the presence of IGWs in the massive 
primary.}

There is much debate concerning the theoretical instability strips of SPB and $\beta$~Cep pulsators. 
Several studies have demonstrated that the edges of the instability strips can change drastically
depending on the choice of metallicity \citep{Daszynska:2013}, opacity table  \citep{Salmon:2012,
Walczak:2015,Moravveji:2016}, and rotation rate \citep{Szewczuk:2017}. One of the most important 
implications of this is that these stars require a sufficiently high (nearly solar) metallicity in 
order to excite pulsations via the $\kappa$-mechanism. Furthermore, \citet{Southworth:2020} have 
demonstrated that in addition to the canonical heat-driven pulsations present in $\beta$~Cep star, 
B-type stars in binaries can also host tidally perturbed oscillations. However, we note that due 
to the low amplitudes expected in such pulsations, we are not likely to observe them with the 
current data set. {To further complicate this picture, both 2D and 3D simulations have 
demonstrated that IGWs can resonantly excite p  and g modes \citep{Edelmann:2019,Ratnasingam:2020}. 
With the current frequency resolution, however, we are not able to distinguish between coherently
driven modes excited via the $\kappa$-mechanism and modes resonantly excited via IGWs.}

{The evolutionary status of the primary is consistent with the instability regions of
both low-order $\beta$~Cep type p and g modes, as well as high-order SPB like g mode pulsations.
\citep{Godart:2017}.}
Depending on the choice of metallicity, opacity enhancement, and / or rotation rate, 
p mode pulsations are only predicted toward the middle to latter half of the main-sequence. 
Based on the evolutionary stage of the primary, the detection of p modes is consistent with 
low {overtone} non-radial p modes. Additionally, both p and g modes are expected to be 
shifted to higher frequencies in the presence of rapid rotation, making the observed g  and 
p mode pulsations consistent with low-{overtone} non-radial oscillations. 

\citet{Burssens:2020} recently reported on the detection of SPB, $\beta$~Cep, and hybrid pulsators
observed in the TESS 2 minute cadence data. Of the 98 stars in their sample, only 9 are eclipsing 
binaries, and only three are hybrid pulsators. Interestingly, none of the EBs are observed to host
hybrid pulsating stars. To date, there are only a handful of hybrid SPB / $\beta$~Cep stars known
in binary systems {\citep[see e.g., ][]{Degroote:2012, Gonzalez:2019} and a handful of either SPB or 
$\beta$~Cep pulsators in systems that are eclipsing, \citep[see e.g., ][]{Freyhammer:2005,Tkachenko:2014,
Jerzykiewicz:2015}.} While there are several known $\beta$~Cep or SPB pulsating stars in clusters 
\citep{Balona:1983,Handler:2007,Saesen:2013,Mozdierski:2019}, no hybrid pulsating SPB/$\beta$~Cep 
stars in EBs have been identified in clusters to date. Although, this is likely a consequence of 
data quality and duty cycle, rather than a physical deficiency.

\subsection{Rotational signal}

In addition to the pulsational variability, {any rotational modulation is also expected
to occur in the low-frequency regime.} To test this, we consider two scenarios: i) rotation
synchronous with the orbital period, i.e. $f_{\rm rot}=0.21759\pm0.00001$~d$^{-1}$ and ii) 
rotation consistent with the projected rotational velocity measured by \citet{Huang:2006b}, 
$v\sin i= 216 \pm 11$~km~s$^{-1}$, i.e. {$f_{\rm rot}=0.88\pm0.05$~d$^{-1}$}, using the 
values for $R_1$ and $i$ from Table~\ref{tab:mcmc}. Both of these values are plotted in 
Fig.~\ref{fig:residuals_ls}, with the rotation rate derived assuming synchronous rotation in 
red and the rotation frequency derived from \citet{Huang:2006b} in grey. There is no significant
signal remaining in the residuals at the orbital frequency. However, $f_4$ in 
Table~\ref{tab:frequencies} agrees with the derived rotation rate from \citet{Huang:2006b}, 
within {$2\sigma$. Assuming $f_{\rm rot}=0.88\pm0.05$~d$^{-1}$, we find} $v/v_{crit}=44 
\pm 4$\%, using the values from Table~\ref{tab:mcmc}. The rotation velocity of the primary 
is consistent with the observed distribution of projected rotational velocities of B-stars 
in clusters \citep{Garmany:2015}. 

{The potential identification of a photometric signal associated with the rotation rate} raises further questions. Given the stably stratified radiative envelope predicted in a 
typical {9.7}~M$_{\odot}$ star, star-spots are not expected. Instead, if a surface 
inhomogeneity is observed, it could be caused by a strong magnetic field that induces 
surface chemical peculiarity \citep{Alecian:2015,DavidUraz:2019}. 

An alternative mechanism 
for inducing photometric variability that coincides with the rotation period is rotationally 
modulated clumpy winds, as has been observed by \citet{Aerts:2018a}. To frame the context 
of HD~149834, only about $\sim10$\% of OB stars are observed to have a detectable surface 
magnetic field. Furthermore, whereas non-magnetic chemically peculiar late B-type stars such 
as HgMn stars are observed to have a high binary incidence \citep{Takeda:2019}, higher mass 
magnetic stars in close binaries are far less common \citep{Alecian:2015}. However, the 
classification of the primary as magnetic or chemically peculiar requires high signal-to-noise
spectra and spectro-polarimetric observations, which are beyond the scope of this paper.

\section{Summary and Conclusions}\label{sec:summary}

In this work, we presented a detailed analysis of the extreme mass ratio eclipsing binary HD~149834.
According to distance and kinematic measurements, the system is a likely member of the young open cluster 
NGC~6193, situated roughly 1~kpc away. Using archival radial velocity and SED measurements combined with
recent TESS photometry, we provide mass and radius estimates for the components of HD~149834, revealing 
the primary to be a {$\sim$9.7$~M_{\odot}$} star. Isochrone fitting demonstrates that the age of the system is 
compatible with independent age estimates for the cluster as well, corroborating its membership. However, 
we do note an age gradient between the most massive cluster members in the core and HD~149834, 
which is consistent with an outside-in star formation scenario. Investigation of the TESS lightcurve 
residuals after removal of the binary signal reveals intrinsic variability consistent with p and g mode 
oscillations, as well as {IGWS}. This makes HD~149834 a rare {moderately rotating} 
hybrid SPB / $\beta$~Cep pulsating star with {IGWs}, in an eclipsing binary belonging to an open 
cluster. 

{The formation of the HD$\,$149834 system---and that of
B-type main-sequence primaries with short-period ($P<10$~d)
extreme mass ratio ($q<0.2$) secondaries in general---may require the secondary
to form at a large separation and to subsequently
migrate inward.
\citet{moe_new_2015} discuss a few plausible pathways,
including early tidal capture, Kozai cycles plus tidal
friction, and fine-tuned accretion conditions in the circumbinary disk.
The eventual evolution and mass transfer that is expected to occur
once the primary evolves off the main sequence could turn the system 
into a low-mass X-ray binary, and may also eventually yield a Type Ia
supernova 
\citep{kiel_populating_2006,ruiter_delay_2011}.}

{The detection of stochastic IGWs and coherent pulsations in the primary is 
consistent with both observations of early B type stars and theoretical predictions
of pulsation modes in such stars.} This system presents a unique opportunity to 
calibrate cluster evolution and star formation models given the independent mass 
and age estimates of the most massive members of the cluster. Additionally, the presence 
of hybrid SPB / $\beta$~Cep pulsations in HD~149834 provides an independent constraint 
on the age and metallicity of the system. The overlap in theoretical instability strips 
for p and g mode pulsations in {9.7}~M$_{\odot}$ stars only occurs in the latter 
half of the main-sequence. Unfortunately, given the frequency resolution provided by 
one sector of TESS observations, we could not search for period spacing patterns, which 
would allow detailed analysis of the internal stellar structure as has been achieved for 
g mode pulsating stars observed by {\it Kepler} \citep{Moravveji:2015,Szewczuk:2018}. 
Future modelling of NGC~6193 should then consider all of the constraints available in order 
to better discriminate between potential models. 

The identification of systems like HD~149834 is becoming increasingly possible given the photometric coverage and 
cadence provided by TESS. Indeed as TESS collects more data of open clusters, more systems like HD~149834 
should be sought out, given the given the opportunities they provide for understanding stellar evolution.  


\acknowledgements
The authors thank S. Wolk for details about the {\it Chandra\/} observations and providing a plot of the light curve.
K.G.S.\ acknowledges partial support from NASA ADAP grant 80NSSC20K0447.
C.J.\ acknowledges that the research leading to these results has received funding from the European 
Research Council (ERC) under the European Union's Horizon 2020 research and 
innovation programme (grant agreement N$^\circ$670519: MAMSIE), and from the 
Research Foundation Flanders (FWO) under grant agreement G0A2917N (BlackGEM).
D.J.S.\ acknowledges funding support from the Eberly Research Fellowship from The Pennsylvania State University Eberly College of Science. The Center for Exoplanets and Habitable Worlds is supported by the Pennsylvania State University, the Eberly College of Science, and the Pennsylvania Space Grant Consortium. 
{The authors are grateful to the anonymous referee for their helpful critiques that helped to improve the manuscript.}

\bibliography{rev2.bib}














\end{document}